\documentclass[prl,aps,twocolumn,amsfonts,showpacs,superscriptaddress,preprintnumbers]{revtex4-1}

\usepackage{graphicx}
\usepackage{xcolor}
\usepackage{rotating}
\usepackage{amsmath,amssymb,graphics}
\usepackage[utf8]{inputenc}

\usepackage[colorlinks=true, urlcolor=blue, linkcolor=blue, citecolor=red, hyperindex=true, linktocpage=true]{hyperref}

\newcommand{\nn}{\nonumber\\}

\def\CN{\mathcal{N}}

\def\CW{\mathcal{W}}
\def\CZ{\mathcal{Z}}

\def\wfr{\mathfrak{w}}

\def\re{\mbox{Re}}
\def\im{\mbox{Im}}

\def\be{\begin{equation}}
\def\ee{\end{equation}}

\def\bea{\begin{eqnarray}}
\def\eea{\end{eqnarray}}

\begin{document}
\preprint{MIT-CTP/5008\;\; OUTP-18-03P}

\title{Transport peak in thermal spectral function of ${\cal N}=4$ supersymmetric \\ Yang-Mills plasma at intermediate coupling}

\author{Jorge Casalderrey-Solana}
\affiliation{Rudolf Peierls Centre for Theoretical Physics,  Clarendon Lab,  Oxford, OX1 3PU, UK}
\affiliation{Departament de F\'\i sica Qu\`antica i Astrof\'\i sica \&  Institut de Ci\`encies del Cosmos (ICC), Universitat de Barcelona, Mart\'{\i}  i Franqu\`es 1, 08028 Barcelona, Spain}
\author{Sa\v{s}o Grozdanov}
\affiliation{Center for Theoretical Physics, MIT, Cambridge, MA 02139, USA}
\author{Andrei O. Starinets}
\affiliation{Rudolf Peierls Centre for Theoretical Physics, Clarendon Lab,  Oxford, OX1 3PU, UK}

\begin{abstract}
We study the structure of thermal spectral function of the stress-energy tensor in ${\cal N}=4$ supersymmetric Yang-Mills theory at intermediate 't Hooft coupling and infinite number of colors. In gauge-string duality, this analysis reduces to the study of classical bulk supergravity with higher-derivative corrections, which correspond to (inverse) coupling corrections on the gauge theory side. We extrapolate the analysis of perturbative leading-order corrections to intermediate coupling by non-perturbatively solving the equations of motion of metric fluctuations dual to the stress-energy tensor at zero spatial momentum. We observe the emergence of a separation of scales in the analytic structure of the thermal correlator associated with two types of characteristic relaxation modes. As a consequence of this separation, the associated spectral function exhibits a narrow structure in the small frequency region which controls the dynamics of transport in the theory and may be described as a transport peak typically found in perturbative, weakly interacting thermal field theories. We compare our results with generic expectations drawn from perturbation theory, where such a structure emerges as a consequence of the existence of quasiparticles. 
\end{abstract}

\maketitle

{\bf Introduction.---}The dynamics of non-Abelian gauge theories at finite temperature is at the heart of a vast variety of physical processes, including the behavior of electroweak and hadronic matter in the Early Universe and the multitude of complicated collective phenomena observed in heavy ion collisions at RHIC \cite{Ackermann:2000tr,Adler:2003kt,Back:2004mh} and LHC \cite{ATLAS:2012at,Chatrchyan:2012ta,Aamodt:2010pa}. Understanding how such dynamical phenomena emerge from the underlying microscopic theory is an important theoretical challenge.
 
Theoretical analysis of non-Abelian plasmas is a complicated task. For static properties, such as pressure and equation of state,  resummed thermal perturbation theory \cite{Laine:2016hma} and  lattice gauge theory  \cite{Borsanyi:2013bia}, respectively, provide a satisfactory description in their complementary domains of applicability, i.e. at weak and strong coupling. Dynamical properties, such as transport coefficients and emission rates,  are, however, not directly accessible to lattice calculations. At weak coupling, these properties can be analyzed within perturbation theory \cite{Arnold:2000dr,Arnold:2003zc,Ghiglieri:2018dib}  but even in this limit, the complicated infrared structure of perturbative diagrams limits the accuracy of those methods. Although different approximation schemes and effective theories have been developed  over the years (see \cite{Laine:2016hma} for a recent review), their applicability to the domain of  intermediate coupling remains limited. To a large extent, our understanding of the dynamics in this region is based on extrapolations from perturbative 
results.

These theoretical limitations provide a strong motivation to study the plasma phase of other non-Abelian gauge theories, for which methods of doing 
calculations at strong coupling exist. In particular, for $SU(N_c)$, $\mathcal{N}=4$ supersymmetric Yang-Mills theory (SYM) with an infinite number of colors, $N_c \to \infty$, the gauge-gravity duality (holography) \cite{Maldacena:1997re} provides a simple, classical computational tool for analyzing its properties in the limit of (infinitely) large `t Hooft coupling $\lambda = g^2_{YM} N_c$. In recent years, the duality was used to obtain new insights into the dynamics of strongly coupled non-Abelian plasmas \cite{CasalderreySolana:2011us}. The duality also allows us to understand corrections to the infinite coupling limit: on the gravity side, they are encoded in the higher-derivative corrections to the Einstein-Hilbert action. Such corrections are necessarily treated as small perturbations of the original second-order equations of motion. Extrapolating results to the regime of finite coupling is subtle, since different physical quantities show different degrees of sensitivity to the corrections \cite{Waeber:2015oka} and adding other higher-order terms may influence the result significantly. Using these higher curvature terms, finite coupling effects of both equilibrium and out-of-equilibrium dynamics of strongly interacting thermal gauge theories have been explored \cite{Gubser:1998nz,Pawelczyk:1998pb,Buchel:2004di,Buchel:2008ae,Steineder:2013ana,Stricker:2013lma,Grozdanov:2014kva,Waeber:2015oka,Grozdanov:2016vgg,Grozdanov:2016fkt,Grozdanov:2016zjj,Andrade:2016rln,DiNunno:2017obv,Casalderrey-Solana:2017zyh,Atashi:2016fai}.

In this Letter, we continue the study of non-perturbative thermal physics at intermediate coupling by providing a stringent test of the consistency of holographic extrapolations. We show how an essential feature of weakly coupled thermal field theories arises from holography---namely, we observe the emergence of a separation of scales in the relaxation of small fluctuations of the plasma at large, but finite coupling, which gives rise to a narrow structure known as the transport peak in the small frequency region of the stress-energy tensor spectral function at zero spatial momentum. The existence of such a structure is a generic expectation for a perturbative plasma and lattice gauge theory. We thus show that holographic extrapolations capture an important qualitative feature of physics at intermediate coupling.

\begin{figure*}[t]
\centering
\includegraphics[width=0.41\textwidth]{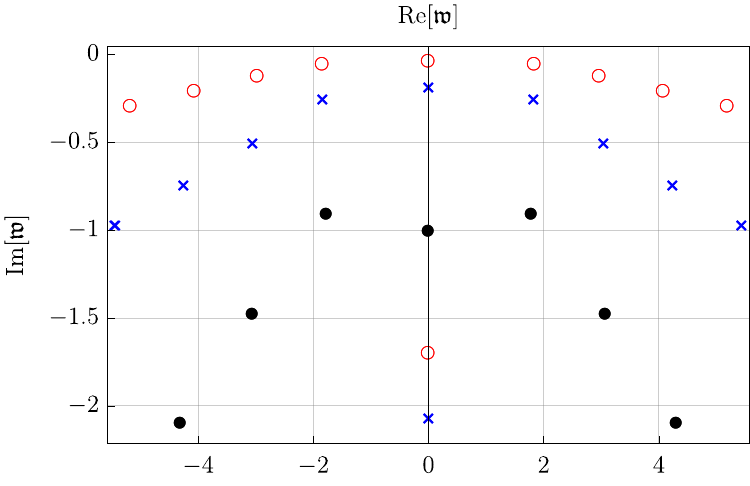}
\hspace{0.08\textwidth}
\includegraphics[width=0.41\textwidth]{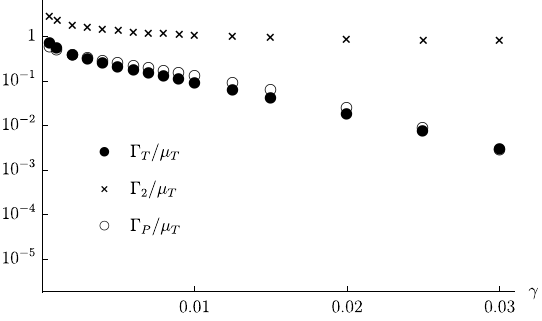}
\caption{Left: Poles of the $\CN=4$ SYM stress-energy tensor correlator $G_{xy,xy}^R (\wfr, 0)$ for  $\gamma = 10^{-3}$ (black solid circles), $\gamma = 10^{-2}$ (blue crosses) and $\gamma = 2 \cdot10^{-2}$ (red empty circles) in the complex plane of normalized frequency $\wfr=\omega/2\pi T$. Right: Imaginary parts of the poles closest to the origin normalized by the  real part of the first complex mode (see Eq.~\eqref{defin-w}) as 
functions of the (inverse) coupling.}
\label{fig:poles}
\end{figure*}

{\bf Gravity dual description at finite coupling.---}The dual description of $\CN=4$ SYM at infinite $N_c$ and large `t Hooft coupling $\lambda$ is 
governed by the low-energy effective action of type IIB supergravity on $AdS_5 \times S^5$. The leading-order finite coupling correction corresponds to the four-derivative term in the action proportional to $\alpha'^3$. The action, upon a reduction on $S^5$, is given by
\begin{align}
S_{IIB} = \frac{N_c^2}{8 \pi^2} \int d^5 x \sqrt{-g} \left(R  - 2 \Lambda + \gamma \CW +\cdots\right)\,.
\label{eq:hd-action}
\end{align}
Here, $\Lambda = - 6/L^2$ is the cosmological constant, $\CW$ is a four-derivative contraction of Weyl tensors (see \cite{Grozdanov:2014kva,Grozdanov:2016vgg}) and $\gamma = \alpha'^3 \zeta(3) / 8$, where in terms of the 't Hooft coupling, $\alpha' / L^2 = \lambda^{-1/2}$. We shall henceforth set $L=1$. The ellipsis in Eq.~\eqref{eq:hd-action} stands for the $O(\gamma)$ corrections to supergravity fields other than the metric: as argued in Ref.~\cite{Buchel:2008ae}, their presence can be ignored for the background under consideration. Also, the equations of motion following from the $10d$ unreduced action and the $5d$ action \eqref{eq:hd-action} are equivalent to $O(\gamma)$, up to a field redefinition \cite{Buchel:2008ae}.

A thermal state of the $\CN=4$ SYM plasma is dual to a black brane geometry in the $5d$ bulk. To leading order in $\gamma$, the black brane metric derived from the action  \eqref{eq:hd-action} is given by \cite{Gubser:1998nz,Pawelczyk:1998pb}
\begin{align}
ds^2 = \frac{r_0^2}{u} \left( - f Z_t dt^2 + dx^2 +dy^2 +dz^2 \right) + Z_u \frac{du^2}{4u^2 f}\,,
\label{eq:corrected_metric}
\end{align}
where $f(u) = 1 - u^2$, $Z_t = 1 - 15\gamma\left(5u^2+5u^4-3 u^6 \right)$, $Z_u = 1 + 15\gamma \left(5u^2 + 5 u^4 - 19 u^6 \right) $ and $r_0$ is the non-extremality parameter related to the Hawking temperature via $T = r_0 (1+15\gamma)/\pi$. 

Retarded stress-energy tensor correlators and spectral functions are computed from the dynamics of linearized metric fluctuations \cite{Son:2002sd,Policastro:2002se,Kovtun:2005ev,Kovtun:2006pf}. At zero spatial momentum, $q=0$, rotational invariance allows us to only consider the metric fluctuation 
component $h_{xy}$, with $x$ and $y$ being two of the spatial coordinates. After introducing the variable $Z_1 = u h_{xy}/ r_0^2$, in Fourier space and to leading order in $\gamma$, the equation of motion has the form \cite{Buchel:2004di,Grozdanov:2016vgg}
\be
\label{eq:fulleom}
Z''_1 - \frac{1+u^2}{u f} Z'_1 + \frac{\wfr^2}{u f^2} Z_1 = \gamma \sum^4_{i=0} \tilde S_{i} (u, \wfr) \, Z_1^{(i)} \, ,
\ee
where $\wfr\equiv\omega/2\pi T$ and $Z_1^{(i)} \equiv \partial^i Z_1 / \partial u^i$. The right-hand side of Eq.~\eqref{eq:fulleom} can be rewritten to include only $Z_1$ and $Z_1'$ at the expense of introducing terms at the order $\mathcal{O}(\gamma^2)$:
 \be
\label{eqn:eqZ}
\!\!\! \sum^4_{i=0} \tilde S_{i} (u,\wfr) \, Z_1^{(i)} = S_0(u,\wfr) Z_1 + S_1(u, \wfr) Z_1' + \, \mathcal{O}(\gamma)   \,, 
\ee
where $S_0$ and $S_1$ can be found in \cite{Grozdanov:2016vgg}.  The resulting ODE is now of second order and can be formally solved (e.g. numerically) for any finite $\gamma$, thus effectively resuming a set of $\gamma$-dependent corrections \cite{Waeber:2015oka}. This procedure is analogous to finding a bound state wave function with a non-trivial electric charge dependence from leading-order potentials in quantum mechanics.

The retarded correlator $G_{xy,xy}^R$ and its associated spectral function $\rho_{xy,xy} (\omega,q) = - 2 \, \text{Im} \, G_{xy,xy}^R (\omega, q)$ are determined from the normalized solutions ${\cal Z}_1(u, \omega)$ to Eqs. \eqref{eq:fulleom}--\eqref{eqn:eqZ}, obeying the incoming wave boundary 
condition at the horizon \cite{Son:2002sd,Grozdanov:2016fkt}: 
\begin{align}
\label{eq:GR}
\rho_{xy, xy}(\omega, q)&= \frac{N_c^2 r_0^4}{2 \pi^2} \lim_{u\to0} \frac{1}{u} \text{Im} \left[ \CZ_1 (u, -\omega)  \CZ_1'  (u, \omega) \right]  .
\end{align}
The relation to the shear viscosity $\eta$ is given by the formula (see e.g. \cite{Baier:2007ix})
\begin{equation}
\rho_{xy, xy}(\omega, 0) = 2  \eta  \omega + O(\omega^3)\,.
\label{Kubo}
\end{equation}
Thus, the shear viscosity is proportional to the value of the function $\rho_{xy, xy}(\omega, 0)/\omega$ at the origin. Here, we compute $\rho_{xy, xy}(\omega, 0)$ numerically and non-perturbatively in $\gamma \sim \lambda^{-3/2}$, which enables us to extrapolate the results to intermediate $\lambda$. 

\begin{figure*}[t]
\centering
\includegraphics[width=0.415\textwidth]{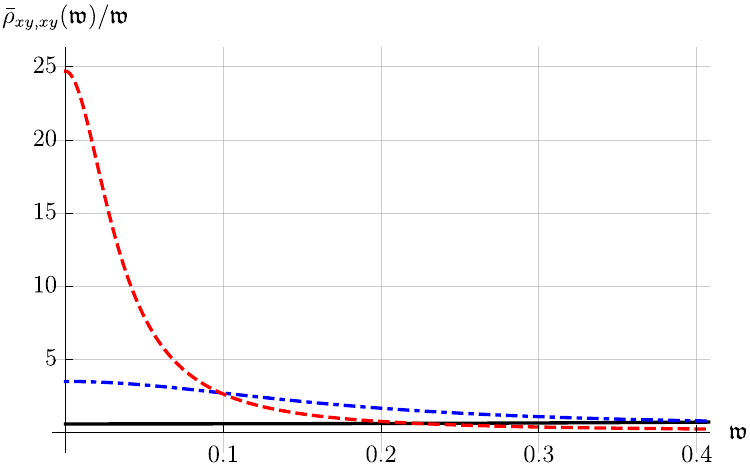}
\hspace{0.08\textwidth}
\includegraphics[width=0.415\textwidth]{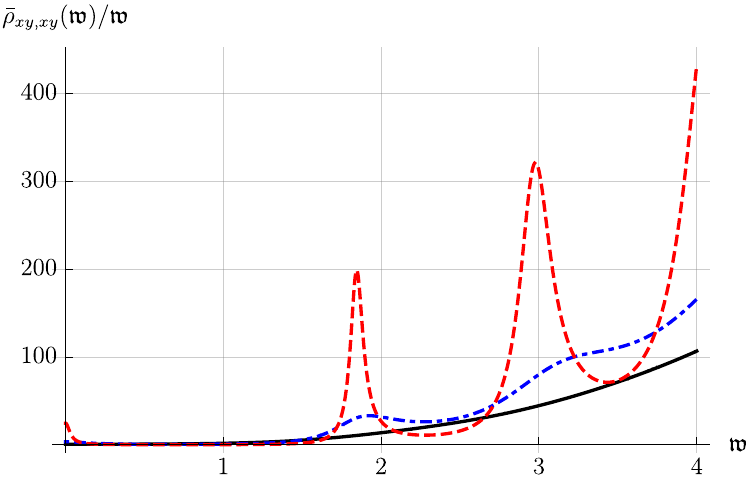}
\caption{Dimensionless spectral function $\bar \rho_{xy,xy} (\omega, q) \equiv  \left( \pi^2 / N_c^2 r_0^4 \right) \rho_{xy,xy} (\omega, q) $  as a function of $\wfr=\omega/2\pi T$ for $\gamma = 10^{-3}$ (black, solid), $\gamma$ = $10^{-2}$ (blue, dot-dashed) and $\gamma$ = $2 \cdot 10^{-2}$ (red, dashed). The left 
and right panels show different frequency ranges.
 }
\label{fig:spectralf}
\end{figure*}

{\bf The transport peak.---}The poles of $G^R_{xy, xy}$ coincide with a discrete set of relaxation modes of the black brane known as quasinormal modes (QNMs) \cite{Son:2002sd,Kovtun:2005ev}. 
The spectrum of these modes in the complex $\wfr$-plane is shown in the left panel of Fig.~\ref{fig:poles} for several values of $\gamma$ at $q=0$.
At infinite coupling ($\gamma = 0$), the black brane QNM spectrum consists of a discrete set of complex frequencies, $\omega^{(i)}_{C}$, $i=1,2,\ldots$, with comparable real and imaginary parts \cite{Starinets:2002br}. At small, but non-vanishing  $\gamma$, the modes $\omega^{(i)}_{C}$ are 
altered perturbatively \cite{Stricker:2013lma,Waeber:2015oka}.  Moreover, one also observes the emergence of a new set of modes, 
$\omega^{(i)}_{I}$, concentrated along the imaginary axis \cite{Grozdanov:2016vgg}. The appearance of these purely dissipative modes has been observed also for other higher-derivative backgrounds, in particular,  for Gauss-Bonnet gravity, where the coupling parameter can be treated non-perturbatively \cite{Grozdanov:2016vgg,Grozdanov:2016fkt}. 

For very small $\gamma$, the modes $\omega^{(i)}_{C}$ closest to the real axis all have $|\omega^{(i)}_C| \ll |\omega^{(i)}_I |$, which is consistent with the fact that the new modes, $\omega^{(i)}_{I}$, decouple from the spectrum as $\gamma \to 0$. As $\gamma$ increases, the distance between neighboring modes in both sets  decreases and they all move toward the real axis. At sufficiently large $\gamma$, a separation of scales emerges in the behavior of the two types of modes, which has important consequences for the structure of the spectral function. To illustrate this separation of scales, we use an appropriate 
Mittag-Leffler expansion for the retarded thermal correlator $G^R (\omega, q=0)$, assumed to be a  meromorphic function in the finite complex frequency plane, obeying the symmetry property 
$G^{R *}(\wfr)= G^R(-\wfr)$ for real $\wfr$ and having the asymptotic behavior $G^R \sim \wfr^4 \ln{\wfr}$ at $|\wfr|\rightarrow \infty$:
\begin{align}
& G^R_{xy, xy} (\wfr, 0) = \frac{i R_{I,1}}{\wfr - \wfr_{I,1}} + \frac{R_{C,1}}{\wfr - \wfr_{C,1}} -  \frac{R_{C,1}^{*}}{\wfr + \wfr_{C,1}^{*}} \nonumber \\
&+ P_4 (\wfr) +  \wfr^5\, \sum\limits_{l>1} \frac{ R_{I,l}}{ |\wfr_{I,l}|^5 (\wfr - \wfr_{I,l})} \nonumber \\ &+
 \wfr^5\sum\limits_{n>1}\Biggl[  \frac{R_{C,n}}{ \wfr_{C,n}^5 (\wfr - \wfr_{C,n})} +  \frac{R_{C,n}^{*}}{\wfr_{C,n}^{*\, 5} (\wfr + \wfr_{C,n}^{*})} \Biggr]\,.
\label{ml-exp}
\end{align}
Here $R_{I,l}$ and  $R_{C,n}$ are the residues at the poles $\wfr= \wfr_{I,l}$ and $\wfr= \wfr_{C,n}$, respectively, $P_4=c_0+i c_1 \wfr + c_2 \wfr^2 + i c_3 \wfr^3 + c_4 \wfr^4$, and $c_i, R_{I,l} \in \mathbb{R}$. The form of the expansion \eqref{ml-exp} was chosen to single out the contributions of the three poles closest to the origin. Let us define
\be
\Gamma_T \equiv -\im\,  \wfr_{I,1}  ,  \quad  \Gamma_P \equiv -\im\,  \wfr_{C,1} ,  \quad \mu_T\equiv \re \, \wfr_{C,1} \,. 
\label{defin-w}
\ee
Then, at $\wfr \ll 1$, we have 
\begin{align}
\label{eq:GRLM}
&\rho_{xy, xy}(\wfr, 0)/\wfr = - \frac{2  R_{I,1}}{\wfr^2 +\Gamma_T^2} - 2 c_1 \nn
&+ \frac{8 \mu_T \Gamma_P \re\, R_{C,1} + 4 (\mu_T^2 - \Gamma^2_P) \im \, R_{C,1}}{(\mu_T^2 + \Gamma_P^2)^2} + O(\wfr^2)\,.
\end{align}
In the limit $\lambda\rightarrow \infty$ ($\gamma \rightarrow 0$), since $\Gamma_T\rightarrow \infty$ and $\re\, \wfr_{C,n}\sim \im\, \wfr_{C,n} \sim n$, the spectral function has no structure around $\wfr = 0$ \cite{Teaney:2006nc,Kovtun:2006pf} and the main contribution to the shear viscosity comes from the coefficient $c_1$ in Eq. \eqref{eq:GRLM}. At finite coupling, however, a characteristic new shape of the spectral function centered around $\wfr=0$ emerges as a consequence of the new pole at
$\wfr  = \wfr_{I,1} = -i \Gamma_T (\gamma)$ moving up the imaginary axis with $\gamma$ increasing. {\it A priori}, the supremacy of this new pole is not guaranteed since all the poles move toward the real axis as $\gamma$ increases, but it turns out that it decouples from the contributions of $\wfr_{C,n}$.  In Fig.~\ref{fig:poles}, we show that the ratios 
$\Gamma_T/\mu_T$ and $\Gamma_P/\mu_T$ are rapidly decreasing functions of $\gamma$. With $\Gamma_T/\mu_T \ll 1$, $\Gamma_P/\mu_T \ll 1$ and the ratio of residues remaining, as we checked, bounded, it is clear that the first term in Eq.~\eqref{eq:GRLM} becomes dominant already for $\gamma \sim 10^{-2}$. In Fig. \ref{fig:poles}, we also show the second purely dissipative mode, $\Gamma_2 \equiv  - \im \, \wfr_{I,2}$, 
which remains comparable to $\mu_T$ for all $\gamma$. This means that its contribution to the spectral function is broad, 
since it overlaps with the contribution of $\wfr_{C,1}$. 

Thus, we find that at large but finite coupling, the low-frequency part of the thermal spectral function is dominated by the contribution of the smallest purely dissipative pole (quasinormal mode). We can therefore identify the two distinct frequency regions in the spectral function
\be
\label{eq:rhoseparated}
\rho_{xy,xy}^{\rm Hol} (\wfr) = \wfr \rho^{\rm Hol}_L (\wfr) + \wfr \rho^{\rm Hol}_H (\wfr) \,,
\ee
where $\rho^{\rm Hol}_L $ comes from $\wfr_{I,1}$ and is dominant in the region $\wfr \lesssim \Gamma_T$, while $ \rho^{\rm Hol}_H $ contains 
contributions from all the remaining poles. For well separated scales, $\rho^{\rm Hol}_L $ is  well approximated by
\be
\label{eq:rhoseparation}
\rho^{\rm Hol}_L \approx \frac{ 4 \pi T \eta \Gamma_T ^2}{\wfr^2 +\Gamma^2_T } \approx \frac{ 2  \pi^2 N_c^2 T^4  \left(1 + 135 \gamma\right) }{ \left(373 \gamma - \ln 2 \right)^2 \wfr^2 + 4 } \,,
\ee
where the shear viscosity $\eta$ is introduced through Eq.~\eqref{Kubo}, and the approximate analytic result on the right hand side follows from the hydrodynamic calculation of $\CZ_1$ to leading order in $\wfr$ and $\gamma$ \cite{Grozdanov:2016vgg}. In this limit, transport is controlled by a single QNM, which leads to a narrow structure in the spectral function---a transport peak. With the above normalization, the area under the low-energy part of $\rho_{xy,xy}^{\rm Hol} (\wfr)/\wfr$ is simply given by $4 \pi T \eta \Gamma_T$. 

Numerically computed spectral functions at $q=0$ (cf. Eq. \eqref{eq:GR}) are shown in Fig. \ref{fig:spectralf}. In the left panel, we plot the low-frequency region, which at very strong coupling (small $\gamma$) shows no characteristic structures for $\omega \ll T$, indicating the absence of  (colored) quasiparticle excitations \cite{Teaney:2006nc}. As the coupling decreases ($\gamma$ increases), a transport peak, associated with the dissipative mode $\omega_I^{(1)}$, emerges. The high-frequency part of the spectral function for different values of $\gamma$  is shown in the right panel of Fig. \ref{fig:spectralf}. For small $\gamma$, the high frequency asymptotics of the spectral function is comparable to that of the $\gamma=0$ limit ($\rho_{xy,xy}^{\rm Hol} (\wfr) \sim \wfr^4$), which for $\omega\gg T$ is fixed by conformal symmetry. At intermediate coupling, however, narrow structures emerge in the region  $\wfr \gtrsim  \mu_T$. These structures reflect the fact that the imaginary parts of the QNMs in the set  $\omega^{(i)}_{C}$ are also small, as already observed in  other channels in Ref. \cite{Grozdanov:2016vgg}. In the limit in which the widths of these structures are much smaller than their real parts, they may be viewed as long-lived bosonic colorless excitations, possessing the same quantum numbers as the stress-energy tensor. 

{\bf From strong to weak coupling.---}We now compare the holographic results for the spectral function extrapolated to the regime of
 finite coupling to the expectations arising from perturbative thermal field theory. For sufficiently large $\gamma$, the separation of scales and the 
 emergence of the transport peak in the low-frequency region of the spectral function implies that transport phenomena occur at much longer time scale, 
 $t_{T } \sim \hbar/\Gamma_{T} k_B T$,  than the typical time scales of other microscopic processes in the plasma, which occur at times $t_{\rm micro}> \hbar/\mu_T k_B T$. This is reminiscent of the dynamics of weakly coupled non-Abelian plasmas, where the hierarchy of energy scales $\epsilon_n \sim \lambda^n T$ emerges for $\lambda \ll 1$. As a consequence of the scattering processes among the plasma constituents, at finite $T$, both bosons and fermions acquire thermal masses squared $\mu_T^2 \equiv m_T^2/T^2 \sim \lambda$ 
and  (momentum-dependent, dimensionless)  thermal widths  \cite{Moore:2008ws} $\Gamma_p\equiv \gamma_T/ T \sim \lambda^2 \ll \mu_T$  for typical momenta, $p\sim T$ (we use the same notations as in holography to emphasize the similarity). Therefore, both bosons and fermions in $\CN=4$ SYM theory can be treated as well-defined colored quasiparticles.

As expected, these dynamical scales manifest themselves in the structure of thermal correlators of gauge-invariant operators. Since the in-medium quasiparticles are colored, they only contribute to the correlator via loop diagrams. For the $q=0$ stress-energy tensor spectral function, a closed-form expression can be obtained in the strict $\lambda=0$ limit. For $\CN=4$ SYM theory, the spectral function is \cite{Teaney:2006nc}
\be
\label{eq:rhopert}
\frac{\rho^{\lambda=0}_{xy,xy} }{ \wfr} = c\,  \delta (\wfr)  + \frac{9 c }{8}\left( \rho_B (\wfr) +  \frac{2}{3} \rho_F (\wfr) \right)\wfr^3,
\ee
where $c = 2 \pi^2 N_c^2 T^4/15$, $ \rho_B=\left(1+ 2 n_B(\pi \wfr)\right)$ and $ \rho_F=\left(1- 2 n_F(\pi \wfr)\right)$, with $n_B$ and $n_F$ being the free Bose and Fermi distributions, respectively. In analogy with Eq.~\eqref{eq:rhoseparated}, the expression \eqref{eq:rhopert} also exhibits two distinct frequency regions: the high-frequency region, proportional to $\wfr^3$, originates from a branch cut associated with the creation of the particle/anti-particle pairs with opposite spatial momentum;  the low-frequency region, proportional to $\delta(\omega)$, arises from the free zero momentum particles. The infinite slope of $\rho_{xy,xy}(\wfr,0)$ at the origin implies that 
at $\lambda=0$, the shear viscosity diverges. 
 
Small coupling corrections to the spectral function preserve the existence of these two distinct structures \cite{Teaney:2006nc}, making the scales at which each of the processes dominates apparent.  Since quasiparticles acquire thermal masses, threshold-like effects occur at $\wfr \sim \mu_T$, separating this contribution from the origin. Because of those effects, ``cusps" are observed in certain correlators in this region \cite{Laine:2011xm}, which may be compared to the narrow structures shown in the holographic computation. However, these structures were not observed in the evaluation of the stress-energy tensor spectral function in \cite{Vuorinen:2015wla}. 

The low-frequency part is more subtle. The contribution proportional to $\delta(\omega)$ emerges as a consequence of a pinching pole in a loop calculation and is the zero coupling manifestation of the transport peak. At finite coupling, the width of the in-medium propagators regularizes this pole and the delta-function acquires a width $\Gamma_T \sim \lambda^2$  \cite{Aarts:2002cc}. This contribution is, therefore, parametrically distinct from the high-frequency region. In perturbation theory, contributions coming from the momenta of order $\Gamma_T T$ suffer from infrared problems which demand resummation of certain classes of diagrams, achieved via an effective kinetic theory  \cite{Arnold:2000dr,Arnold:2003zc}. The analysis of the spectral function using the effective kinetic theory has been performed in \cite{Petreczky:2005nh,Moore:2006qn,Moore:2008ws,Hong:2010at,Moore:2018mma}. In particular, in Ref.~\cite{Moore:2018mma} it was argued that in a scalar finite-temperature field theory, the transport peak in $\rho_{xy,xy} (\wfr)/\wfr$ at weak coupling arises as a consequence of a branch cut along the imaginary frequency axis in the expression for the  retarded correlator, with non-uniform discontinuity along the cut peaked at $\im\, \wfr \sim \lambda^2$. It is also known that for current-current correlators in QCD, the low-frequency region is well approximated by the Lorentzians as the coupling increases \cite{Moore:2006qn}, which is consistent with our holographic extrapolation.

{\bf Discussion.---}The qualitative similarities between our holographic extrapolation and perturbation theory expectations are quite remarkable. Nevertheless, we would like to stress that the extrapolation we used does not capture the full contribution from higher orders of the $\gamma$ expansion. While both Eqs. \eqref{eqn:eqZ} and \eqref{eq:GR} are accurate to leading order in $\gamma$, it is unclear whether the corrections resummed by non-perturbative solutions of Eq.~\eqref{eqn:eqZ} are the most relevant ones. Furthermore, since these corrections are induced by higher-derivative terms, their magnitude depends on the momentum scale, apparently making them larger as $\wfr$ increases. For this reason, we expect our extrapolation to better describe the low-frequency dynamics of the plasma, while the high-frequency part, including the structures at $\wfr \sim \mu_T,$ may be more sensitive to other effects. While further analysis of the different corrections is needed, in the absence of  higher-derivative supergravity corrections at higher orders in $\alpha'$, the recovery of many qualitative features of a weakly coupled plasma suggests that this type of extrapolation captures essential aspects of the full result. 

One of those qualitative aspects is the observation of a well-defined transport peak. Based on the analysis of transport, this structure may be taken as an indication of the existence of quasiparticles in the holographic calculation. However, even if the narrow structures observed at finite frequency in Fig. \ref{fig:spectralf} correspond to approximate states in the thermal ensemble in the large $N_c$ limit,  they would not contribute to the leading $N_c$ correlator since those excitations must be colorless. In contrast, in the perturbative computation, both bosonic and fermionic quasiparticles are in the adjoint representation, and they contribute to the connected correlator to order $N^2_c$. It is remarkable that the holographic
calculation, without any explicit reference to quasiparticles, captures these distinct features of the stress-energy tensor correlator. It would be interesting to explore whether, as predicted by the quasiparticle picture, a transport peak occurs at the same parametric scale $\Gamma_T$ in the spectral functions of other conserved currents.  

Finally, we would like to stress that even after the extrapolation to intermediate coupling, the retarded correlator in the holographic calculation remains a meromorphic function. Although the precise analytic structure of this correlator at small non-zero coupling remains unknown, the expectation in perturbation theory \cite{Hartnoll:2005ju,Romatschke:2015gic,Kurkela:2017xis,Moore:2018mma,Grozdanov:2018atb} is that the retarded functions develop branch cut singularities in the complex frequency plane. It would be interesting to further understand how these singularities could emerge in holography, beyond what was observed  in Ref.~\cite{Grozdanov:2016vgg}, as well as in non-conformal plasmas \cite{Russo:2013qaa}.

{\bf Acknowledgements.---}We would like to thank G. Aarts, N. Gushterov, A. Kurkela, G. Moore,  P. K. Kovtun, D. T. Son, A. Vuorinen  and L. Yaffe for useful discussions. J. C. S. is a University Research Fellow of the Royal Society (on leave) and is supported by  grants SGR-2017-754, FPA2016-76005-C2-1-P and MDM-2014-0367.  S. G. was supported by the U.S. Department of Energy under grant Contract Number DE-SC0011090. A. O. S. was supported by the European Research Council under the European Union's Seventh Framework Programme (ERC Grant agreement 307955). 

\bibliographystyle{apsrev4-1}
\bibliography{Genbib}{}

\end{document}